\documentclass[12pt]{article}

\usepackage{epsf,graphicx}

\newcommand{\beq}{\begin{equation}}
\newcommand{\eeq}{\end{equation}}
\newcommand{\beqa}{\begin{eqnarray}}
\newcommand{\eeqa}{\end{eqnarray}}

\def\la {\lambda }

\def\rar{\rightarrow }
\def\be{\begin{equation}}

\def\j-{\J_-}

\def\ee{\end{equation}}

\def\bearr{\begin{eqnarray}}
\def\bearrs{\begin{eqnarray*}}
\def\eearr{\end{eqnarray}}
\def\eearrs{\end{eqnarray*}}
\def\barr{\begin{array}}
\def\earr{\end{array}}

\def\th{\theta}

\def\non\non{\nonumber}
\def\nn8{\nonumber\\[15pt]}
\def\l{\left}
\def\r{\right}

\def\f{\frac}

\def\dis{\displaystyle}

\begin{document}
\begin{flushright}
SINP/TNP/01-19
\end{flushright}
\begin{center}
{\large{\bf Energy deposition due to neutrino pair annhilation near
rotating neutron stars}}
\vskip 10pt
{\it A.R. Prasanna $^{a}$ and Srubabati Goswami$^{b}$ \\
$^{a}$Physical Research Laboratory, Ahmedabad 380009, India}\\
$^{b}$ 
{\it Saha Institute of Nuclear Physics,\\1/AF, Bidhannagar,
Calcutta 700064, INDIA.}\\

\bf{Abstract}
\end{center}
General relativistic effects have been shown to increase the
energy deposition rate due to the process $\nu \bar{\nu} \rightarrow 
e^{+}e^{-}$ in supernovae and neutron stars. 
In this paper we study the effect of inclusion of the rotation
of the star in the general relativistic treatment. 
We show that inclusion of rotation results in a reduction 
in the  heating rate as compared to the no rotation case.

\vskip 20pt

key words: neutrinos$--$supernovae$--$neutron stars 
\vskip 20pt
\section{Introduction}
The  neutrino pair annilation reaction ($\nu \bar{\nu} \rightarrow e^{+}e^{-}
$) is one of the important processes in understanding the energy transfer 
from a hot proto neutron star to the outer layers of a 
supernova \cite{goodman}. 
It was pointed out in \cite{wm89,bethe} that this process can give a 
great boost to the delayed neutrino heating mechanism of Bethe and Wilson
\cite{bw85} in addition to the $\nu$-nucleon capture reactions.
The annihilation process is useful as this can continue giving energy to the 
"radiation bubble" till there is an emitted neutrino flux \cite{bethe}. 
Apart from type II supernovae
this process is also important for collapsing neutron stars
\cite{collnu}, close neutron star binaries in their last stable 
orbit \cite{mw97}.  
In particular, this process has been considered to be one of the 
possible sources 
of energy for gamma-ray bursts \cite{gamma}. 
The reaction efficiency of this process was calculated in \cite{goodman,cooper,
brezinsky} using Newtonian gravity. However strong gravitational effect in 
supernova and collapsing neutron star envioronments 
render a general relativistic 
calculation necessary 
\cite{jaro,janka,salmon,asano,sw}. 
It has been shown in \cite{salmon}  
that the energy deposition rate increases by about a factor of 4 in 
supernovae and by a factor of 30 in neutron stars if general relativistic 
effects are taken into account. In this paper  
we include the effect of rotation in the general relativistic
treatment of \cite{salmon} and  
compare
how much this changes the  reaction efficiency  of the neutrino pair 
annhilation process over the case discussed in \cite{salmon}.

\section{Geodesics in the External Field of a Slowly Rotating Object}
The metric outside a slowly rotating star as given by the
approximate Hartle-Thorne solution, with only the dipole
corrections on a static star is given by \cite{hartle}
\be
\barr{lll} 
ds^2 &=& - \l( 1 - \f{2m}{r} + \f{2J^2}{r^4} \r) dt^2 +
\l( 1 - \f{2M}{r} + \f{2J^2}{r^4} \r)^{-1} dr^2 + r^2
d\th^2\\[10pt] &&+ r^2\sin^2\th \l( d\phi - \f{2J}{r^3} dt\r)^2
\earr 
\ee 
where $M$ is the mass and $J$ the specific angular
momentum. 
Here and through out the paper (unless otherwise mentioned) we use 
the geometrised units (G=c=1). 
Using the Lagrangian appropriate to the motion in the
equatorial plane $\l( \th = \pi /2 \r)$
\be
\barr{lll}
2{\cal L} &=& - \l( 1 - \f{2M}{r} - \f{2J^2}{r^4} \r) \dot{t}^2
+ \l( 1 - \f{2M}{r} + \f{2J^2}{r^4} \r)^{-1} \dot{r}^2\\[10pt]
&& + r^2 \dot{\phi}^2 - \f{4J}{r} \dot{\phi}\dot{t}
\earr
\ee
the generalised momenta may be written as
\be
\barr{lll}
p_t&=& - \l( 1 - \f{2M}{r} - \f{2J^2}{r^4} \r) \dot{t} -
\f{2J}{r} \dot{\phi} = - E\\[10pt]
p_\phi&=& - \f{2J\dot{t}}{r} + r^2\dot{\phi} = L\\[10pt]
p_r&=& \l( 1 - \f{2M}{r} + \f{2J^2}{r^4} \r)^{-1} \dot{r}
\earr
\ee
and thereby the Hamiltonian
\be
2{\cal H} = - E\dot{t} + L \dot{\phi} + \l( 1 - \f{2M}{r} +
\f{2J^2}{r^4} \r)^{-1} \dot{r}^2 = \delta_1
\ee
wherein the constant $\delta_1 = 1$ for time-like geodesics and
$\delta_1 = 0$ for null geodesics \cite{chandra}.

Solving for $\dot{t}$ and $\dot{\phi}$, one gets
\be
\barr{lll} U^t = \dot{t} &=& \l( E - \f{2JL}{r^3} \r) \l( 1 -
\f{2M}{r} + \f{2J^2}{r^4} \r)^{-1}\\[10pt] U^\phi =
\dot{\phi}&=&\l[ \l( 1 - \f{2M}{r} - \f{2J^2}{r^4} \r) \f{L}{r^2}
+ \f{2JE}{r^3} \r] \l( 1 - \f{2M}{r} + \f{2J^2}{r^4} \r)^{-1}
\earr 
\ee 
As we will be interested in zero rest mass particles we
consider the null geodesics $\delta_1=0$, and thereby obtain
$\dot{r}$ from (4) to get the equation
\be
\barr{lll}
 \l( \f{1}{r^2} \f{dr}{d\phi}\r)^2& = &\f{\l( 1-2M/r +
2J^2/r^4\r)^2}{\l( 1 - 2M/r + 2JE/Lr - 2J^2/r^4\r)^2}\\[10pt] &&
\l[ \f{E^2}{L^2} - \f{4JE}{Lr^3} - \f{1}{r^2} \l( 1 - \f{2M}{r} -
\f{2J^2}{r^4} \r) \r] \earr \ee which in the limit $J \rar 0$
reduces to the well-known Schwarzschild form \cite{misner}
\be
\l( \f{1}{r^2} \f{dr}{d\phi}\r)^2 = \f{1}{b^2} - \f{1}{r^2} \l(
1 - \f{2M}{r} \r)
\ee
where $L/E \rightarrow b$, 
the impact parameter for a massless particle.
\\

Introducing the Local Lorentz tetrad $\la^{(a)}_{\;\;\;\; i}$ as
given by
\be
\la^{(a)}_{\;\;\;\; i} = \barr{cccc} \l( 1 - 2M/r + 2J^2/r^4
\r)^{1/2}&0&0&0\\[10pt] 0&\l( 1 - \f{2M}{r} + \f{2J^2}{r^4}
\r)^{-1/2}&0&0\\[10pt] 0&0&r&0\\[10pt] -2J/r^2&0&0&r\sin\th \earr
\ee (upper index refers to rows and the lower to columns),
one can define the angle $\th_r$ between the trajectory and the
tangent vector in terms of local radial and longitudinal velocities
\be
\barr{lll} 
tan \th_r = \f{V^1}{V^3}&=& \f{\l( \la^1_r V^r\r)}{\l(
\la^3_\phi V^\phi + \la^3_{\;\; t}\r)}\\[10pt] &=& \f{\l( 1 -
\f{2M}{r} + \f{2J^2}{r^4} \r)^{-1/2} V^r}{\l( rV^\phi -
\f{2J}{r^2} \r)}\\[10pt] &=& \l( 1 - \f{2M}{r} + \f{2J^2}{r^4}
\r)^{-1/2} \l( r - \f{2J}{r^2V^\phi} \r)^{-1} \l( \f{dr}{d\phi}\r)
\label{thetar}
\earr \ee
As the local velocity $V^\phi = U^\phi /U^t$,  using
(5), one can get
\be
\l( \f{dr}{d\phi}\r) = \l( 1 - \f{2M}{r} + \f{2J^2}{r^4} \r)^{3/2}
\l[ \l( 1 - \f{2M}{r} - \f{2J^2}{r^4} \r) \f{1}{r} +
\f{2J}{br^2} \r]^{-1} \tan\th_{r}
\ee
Eliminating $\l( \f{dr}{d\phi} \r)$ between (6) and (10) and
simplifying, one gets the impact parameter $b$ to be
\be
b = \l[ \f{2J}{r^3} + \f{\l( 1 - 2M/r +
2J^2/r^4\r)^{1/2}}{r\cos\th_r} \r]^{-1}
\ee
For a neutrino emitted tangentially $\l( \th_R = 0 \r)$ from a
neutrinosphere at $R$
\be
\cos\th_r = \f{R^3r^2 \l( 1- \f{2M}{r} + \f{2J^2}{r^4}
\r)^{1/2}}{2J \l( r^3-R^3\r) + R^2r^3 \l( 1 - \f{2M}{R} +
\f{2J^2}{R^4} \r)^{1/2}}
\label{costhr}
\ee
which reduces to 
\[
\cos\th_r = \f{R}{r} \sqrt{\f{1-2M/r}{1-2M/R}}
\]
for $J=0$ \cite{salmon}. 
The minimum
photosphere radius for the non-rotating case was at $R=3M$.
For the present
case the minimum radius  and the corresponding impact parameter 
may be found from the roots of the system of equations 
\be
R^4 - \l( 3 - 6J/b\r) R^3 - 6J^2 = 0
\label{rt1}
\ee
and 
\be
R^6 - b^2 R^4 + 2b(b-2J)R^3 - 2 b^2 J^2=0
\label{rt2}
\ee
obtained from the effective potential for the particle in circular orbit. 
In the above equations we have expressed all the quantitites in units 
of stellar mass M. 
The maximum value of the rotation parameter that we take is 
$J/M^2$ = 0.3 as the metric used is valid for slowly rotating 
configurations. 
It may indeed be verified that this value $J/M^2 (=0.3)$  
is about the critical value for the case of a millisecond pulsar \cite{anshu}. 
In  Table 1 we give the values of  
neutrinosphere radius
 ($R_{\nu}$) ($\sim$ last photon orbit) and the  
impact parameter ($b$) for different values of $J/M^2$ by solving 
simultaneously the eqs. (\ref{rt1}) and (\ref{rt2}). 
For every value of $J/M^2$ there exists
atleast one real root which is always located at a distance $<
3M$, giving the neutrinosphere radius to be less than that in
the case of Schwarzschild solution (the non-rotating case). 
In Table 1 we also give the corresponding event horizons 
($R_{EH}$), obtained by solving the equation 
$g_{tt} =0$, for various values of $J/M^2$. 
In the following section we calculate the energy deposition rates 
at a distance $r>R_{\nu}$.  

\begin{description}
\item Table 1: 
The values of the event horizon ($R_{EH}$), 
neutrinosphere radius
($R_{\nu}$) and the impact parameter (b) for different values of the 
rotation parameter $J/M^2$. 
\begin{center}
\begin{tabular}{||c|r|r|r|r|}
\hline\hline $J/M^2$&$R_{EH}/M$& $R_{\nu}/M$&b/M\\ \hline
0.1&2.00249&2.8771&4.9855\\ \hline
0.2&2.00985&2.7355&4.7483 \\ \hline
0.3&2.02178&2.56546&4.4715 \\ \hline
\hline
\end{tabular}
\end{center}
\end{description}

\section{Energy Deposition Rates} 
The energy depostion rate/unit time/unit volume  by this process
is given in general by \cite{cooper}
\begin{equation}
\dot{q} = \frac{7 D G_F^2 \pi^3 \zeta(5)}{2 c^5 h^6} (kT(r))^9 \Theta(r)
\label{qdot}
\end{equation}
where $G_F$ is the Fermi coupling constant,
and 
$D = 1 \pm 4 sin^2\theta_W + 8 sin^4 \theta_W
$
with $sin^2 \theta_W = 0.23$ and the $+$ sign is for electron type neutrinos
and antineutrinos and the $-$ sign is for the muon and tau type neutrinos
and antineutrinos.
$T(r)$ is the temperature measured by the local observer and
$\Theta(r)$ is the angular integration factor.
A general relativistic treatment requires the incorporation of 
gravitational red shift in $T(r)$ and the effect of bending of the path 
of neutrinos in $\Theta(r)$.
In terms of the unit direction vector 
${\bf{\Omega}}_{\nu}$ and the solid angle subtended $d\Omega_{\nu}
$, $\Theta(r)$ can be written as 
\begin{eqnarray} 
\Theta(r) &=&
 {\bf \int \int} (1 - {\bf {\Omega}_\nu}.{\bf \Omega}_{\bar{\nu}})^2
d{\bf\Omega}_\nu d{\bf\Omega}_{\bar{\nu}} \label{Thetar}\\
&=& 4 \pi^2 {\int_{x}^1}{\int_{x}^1}
[ 1 - 2 \mu_\nu \mu_{\bar{\nu}}
+ \mu_{\nu}^2 \mu_{\bar{\nu}}^2 + \frac{1}{2}
( 1 - {\mu_\nu}^2)( 1 - \mu_{\bar\nu}^2)] d\mu_\nu d\mu_{\bar{nu}}
\label{thetar2}
\end{eqnarray}
where $\mu = sin\theta$, $\Omega = (\mu, \sqrt{1 - \mu^2} cos\phi,
\sqrt{1 - \mu^2}sin\phi )$ and $d{\Omega} = cos\theta d\theta d\phi$.
This simplifies to
\beq
\Theta(r) = \frac{2 \pi^2}{3} ( 1 - x)^4 (x^2 + 4 x + 5)
\eeq
where $x=sin\theta_r$. 
With $\theta_r$ defined in eq.(\ref{thetar}) of the previous section
one gets
\be
x
= \l[ 1 - 
\f{R^6r^4 \l( 1 - \f{2M}{r} + \f{2J^2}{r^4}\r)}
{\l\{ 2J \l( r^3 - R^3 \r) + R^2 r^3 \l( 1 -
\f{2M}{R} + \f{2J^2}{R^4} \r)^{1/2} \r\}^2} \r]^{1/2}
\ee
This reduces to the expression given in \cite{salmon} for $J=0$. 

The neutrino temperature varies linearly with redshift and 
$T(r)$ in eq.(\ref{qdot}) at a radius $r$ is 
related to the neutrino temperature at the neutrinosphere 
radius $R$ as 
\be
T(r) = \sqrt{\frac{1 - \f{2M}{R} - \f{2J^2}{R^4}}
{1 - \f{2M}{r} - \f{2J^2}{r^4}}} T(R)
\label{temp}
\ee 

The total amont of local
energy deposited by $\nu \bar{\nu} \rightarrow e^+e^-$ for a
single neutrino flavour for a rotating star can be defined as 
(in analogy to the non-rotating case discussed in \cite{salmon})  
\beq 
\dot{Q} =
\int_{R}^\infty{\dot{q}} 
\frac{4 \pi r^2 dr} 
{\sqrt{1 - \frac{2M}{r} - \f{2J^2}{r^4}}} 
\eeq 
where ${\dot{q}}$ is defined in eq.(\ref{qdot}) with $T(r)$ and $\Theta(r)$
defined in eqs. (\ref{temp}) and (\ref{Thetar}) respectively. 
This can be simplified to obtain 
\beq 
\dot{Q}_{51} = 1.09 \times 10^{-5}
{\cal{F}}\left(\frac{M}{R}, \frac{J}{R^2} \right) 
D L^{9/4}_{51} R^{-3/2}_6 \eeq
 where $\dot{Q}_{51}$ and
$L_{51}$(luminosity) are in units of $10^{51}$ ergs/sec, $R_{6}$
is the radius in units of 10 km and 

\[
\barr{lll} {\cal F} \l( \f{M}{R}, \f{J}{R^2} \r) &=& 3 \l( 1 -
\f{2M}{R} - \f{2J^2}{R^4} \r)^{9/4}\\[10pt] &&{\dis\int^\infty_1}
\l\{ \l( x -1\r)^4 \l( x^2 + 4x + 5 \r) \cdot \f{y^2dy}{\l(
1 - \f{2M}{r} - \f{2J^2}{r^4}\r)^5}\r\} \earr
\]
where $y = r/R$. 
For $J \rightarrow 0$ this reduces to ${\cal{F}}(\f{M}{r})$ of \cite{salmon}
and the Newtonian limit is obtained by taking $M \longrightarrow$ 0
\cite{cooper}. 

\section{Results and Discussions}

In fig. 1a and fig. 1b  we plot the ratio 
$\dot{Q}(\f{M}{R},\f{J^2}{R})/\dot{Q}_{\rm Newt}$
= ${\cal{F}}(\f{M}{R},\f{J^2}{R})$ vs $R/M$. 
In fig. 1a we plot the ratio from $R/M=5$ to $R/M = 10$ which is relevant for a type II supernova while in fig. 1b we plot the ratio at smaller values of $R/M$
which is relevant for collapsing neutron stars. 
The $J \rightarrow 0$ curves correspond to the case  
without rotation for which  the general relativistic effects enhances the 
energy deposition rate by a factor of 4 at $R=5M$ \cite{salmon} and 
almost by a factor of 30 for $R=3M$. 
As one includes
the rotation of the star there is a drop in the energy deposition rate.  
The reduction is more for smaller
values of $R/M$  and higher values of $J/M^2$. 
If we take $J/M^2 = 0.3$ then at
$R/M = 3$ we get a reduction by $\approx$ 38 \% whereas for $R/M =5$ 
the reduction due to rotation is $\approx$ 9\%. 
With rotation the apparent angular size of the star 
seen by the neutrino decreases (see eq. (\ref{costhr}) 
thus decreasing the probability of head on collision. This results in a
drop in the heating rate. 

In fig. 2 we plot $d\dot{Q}/dr$ vs the radius for three values of R/M 
(5,7 and 10). For each case we plot $d\dot{Q}/dr$ for various values of the 
rotation parameter $J/M^2$. 
We also plot the Newtonian case for which M=0 and J=0.
The heating rate is seen to fall sharply as the radius increases.  
The inclusion of general 
relativistic effects enhances the heating rate   
but introduction of rotation reduces this. 
The reduction is more pronounced for smaller values of R/M and higher 
values of $J/M^2$.
The resultant heating rate after including the rotation is however still
higher than the Newtonian value.  

To conclude, in this paper we have extended the general relativistic calculations of the neutrino heating rates due to neutrino pair annihilation to include 
the rotation of the star. 
We find that the effect of including rotation is to 
reduce the heating rate over the no rotation case and 
the reduction can be as large as 38\%.   
The effect is more pronounced for smaller values of the ratio $R/M$, 
in the 
range relevant for collapsing neutron stars. 
\\

S.G. would like to acknowledge the hospitality extended to her by the theory 
group of Physical Research Laboratory, where some part of the work was 
done.

\vskip 20pt
\centerline {\bf Figure Caption} 

\vskip 10pt
Fig. 1. The ratio of the general relativistic energy deposition
rate to the Newtonian rate for various values of the rotation
parameter.  The solid line is for no rotation, the big dashed line is for
$J/M^2$ = 0.1, the small dashed line is for $J/M^2$ = 0.2 and the
dotted line is for $J/M^2$ = 0.3. \\ \\  \\
Fig. 2. $d\dot{Q}/dr$ as a function of radius for different values 
of $M/R$ and $J/M^2$. Also shown is the variation for the Newtonian
case (M=0,J=0).

\begin{figure}
\centerline{{\epsfxsize=1.3\textwidth\epsfbox{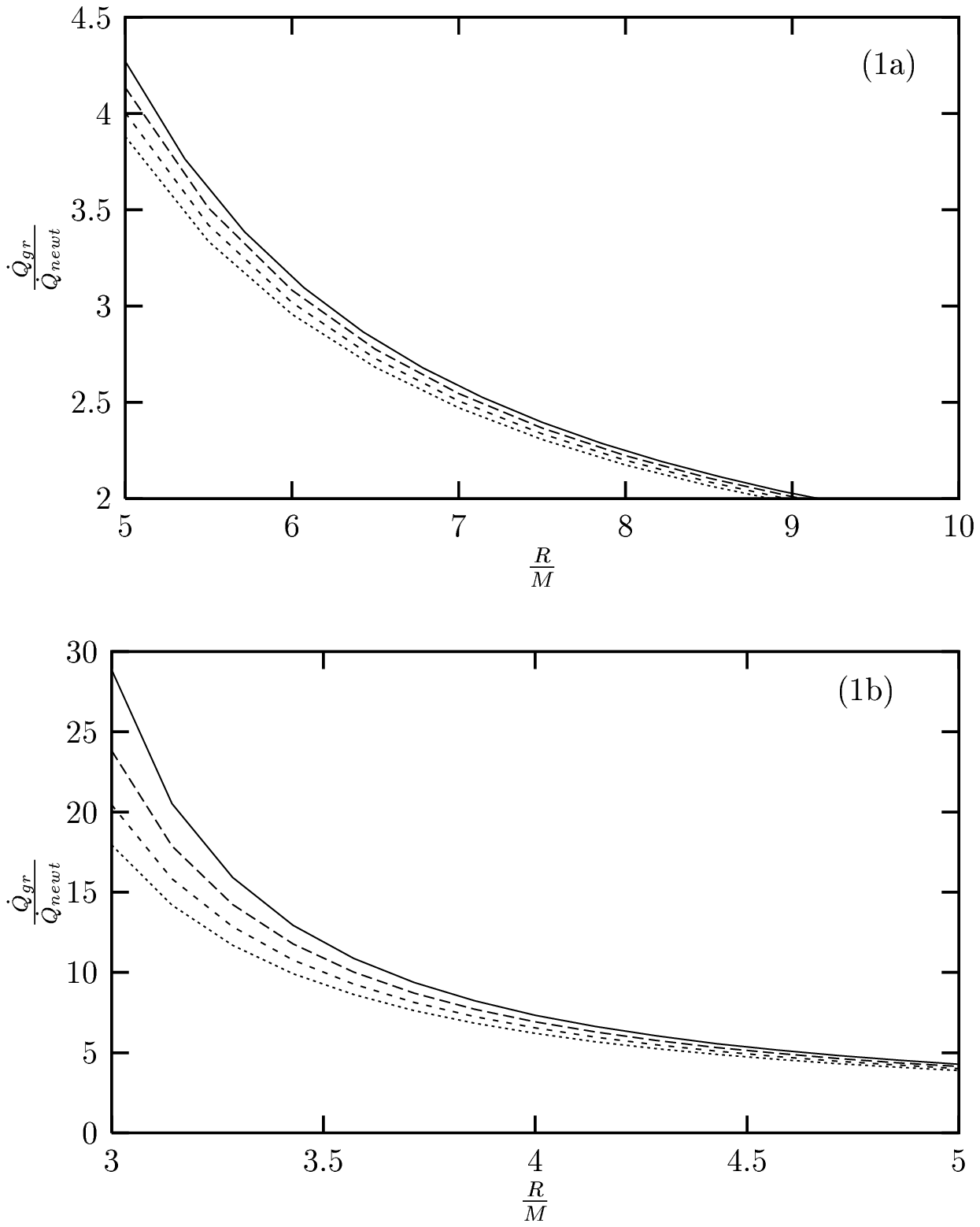}}}
\end{figure}

\topmargin-1in
\begin{figure}
\centerline{{\epsfxsize=.9\textwidth\epsfbox{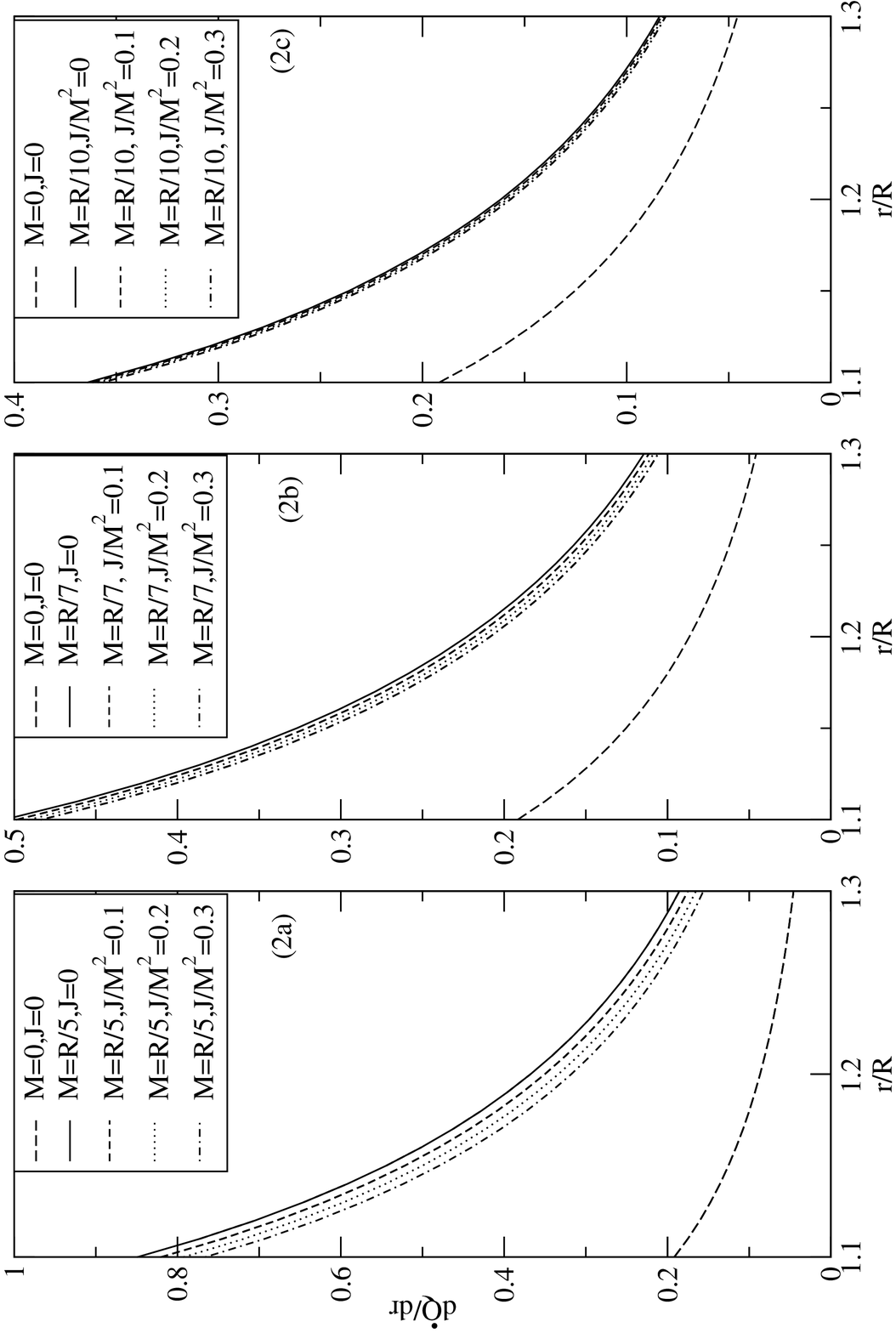}}}
\end{figure}

\end{document}